\documentclass[%
 reprint,
 amsmath,amssymb,
 aps,
floatfix,
]{revtex4-2}

\usepackage{graphicx}
\usepackage{dcolumn}
\usepackage{bm}
\usepackage{xcolor}
\usepackage{ulem}
\usepackage[caption=false]{subfig}

\usepackage[hidelinks]{hyperref}

\def\VEV#1{{\left\langle #1 \right \rangle}}

\begin{document}

\preprint{}

\title{Uncorrelated compensated isocurvature perturbations from kSZ tomography}

\author{Neha Anil Kumar}
\email{nanilku1@jhu.edu}
\affiliation{%
 William H.\ Miller III Department of Physics and Astronomy, Johns Hopkins University
 Baltimore, Maryland 21218, USA
}%

\author{Selim C. Hotinli}
\affiliation{%
 William H.\ Miller III Department of Physics and Astronomy, Johns Hopkins University
 Baltimore, Maryland 21218, USA
}%

\author{Marc Kamionkowski}
\affiliation{%
 William H.\ Miller III Department of Physics and Astronomy, Johns Hopkins University
 Baltimore, Maryland 21218, USA
}%



\begin{abstract}
Compensated isocurvature perturbations (CIPs) are relative density perturbations in which a baryon-density fluctuation is accompanied by a dark matter density fluctuation such that the total-matter density is unperturbed. These fluctuations can be produced primordially if multiple fields are present during inflation, and therefore they can be used to differentiate between different models for the early Universe. Kinetic Sunyaev-Zeldovich (kSZ) tomography allows for the reconstruction of the radial-velocity field of matter as a function of redshift. This technique can be used to reconstruct the total-matter-overdensity field, independent of the galaxy-density field obtained from large-scale galaxy surveys. We leverage the ability to measure the galaxy- and matter-overdensity fields independently to construct a minimum-variance estimator for the primordial CIP amplitude, based on a mode-by-mode comparison of the two measurements. We forecast that a configuration corresponding to CMB-S4 and VRO will be able to detect (at $2\sigma$) a CIP amplitude $A$ (for a scale-invariant power spectrum) as small as $A\simeq 5\times 10^{-9}$. Similarly, a configuration corresponding to SO and DESI will be sensitive to a CIP amplitude $A\simeq 1\times 10^{-7}$. These values are to be compared to current constraints $A \leq {\cal O}(0.01)$.
\end{abstract}

\maketitle


\section{\label{sec:Introduction}Introduction\protect}

Improving our understanding of the statistical characteristics of the primordial density fluctuations of our Universe is one of the primary goals of upcoming large-scale structure surveys and cosmic microwave background (CMB) experiments. The current observations of the small-amplitude $[{\cal O}(10^{-5})]$ temperature and polarization fluctuations in the CMB are consistent with Gaussian adiabatic fluctuations, as predicted by single-field models of inflation. Nevertheless, the search for small deviations from adiabaticity or Gaussianity remains a promising direction of research that can allow us to effectively distinguish between different models of inflation and determine the number of degrees of freedom governing the dynamics of the early Universe ~\citep[e.g.,][]{Baumann:2011nk, Assassi:2012zq, Chen:2012ge, Noumi:2012vr, Arkani-Hamed:2015bza, Lee:2016vti, Kumar:2017ecc, An:2017hlx, An:2017rwo, Baumann:2017jvh, Kumar:2018jxz, Anninos:2019nib, Gong:2013sma, Pi:2012gf}.

One such deviation that is particularly difficult to probe with CMB data alone is the class of isocurvature perturbations that leave the total-matter density unchanged \cite{Gordon:2002gv,Gordon:2009wx,Holder:2009gd}. These compensated isocurvature perturbations (CIPs) may arise in various models of inflation with multiple fields \cite{Linde:1996gt,Sasaki:2006kq,Gordon:2009wx,Lyth:2002my,Gordon:2002gv,Langlois:2000ar,He:2015msa} and also during baryogenesis \cite{DeSimone:2016ofp}. In the multfield models, the CIP fluctuations may be fully correlated with the adiabatic perturbation, completely uncorrelated, or (most generally) somewhere in between. Specifically, uncorrelated CIPs are a characteristic of the baryogenesis model \cite{DeSimone:2016ofp}.

Because CIPs leave the total matter distribution unchanged, they give rise to no CMB fluctuations at linear order. Instead, they induce higher-order effects on the CMB power spectrum \cite{Munoz:2015fdv,Heinrich:2016gqe,Smith:2017ndr,Valiviita:2017fbx,Planck:2018jri}, and the CMB trispectrum \cite{Grin:2011tf,Grin:2011nk,Grin:2013uya}. On small-distance scales, the effects of CIPs may be manifest in CMB spectral distortions \cite{Haga:2018pdl,Chluba:2013dna} or the recombination history \cite{Lee:2021bmn}. Because these higher-order effects are harder to measure, CIPs are rather poorly constrained by the CMB data, with the recent constraints allowing for fairly large-amplitude perturbations. However, there are various other prospects to probe different models of CIPs. For example, the effects of CIPs on baryon acoustic oscillations have been studied in Ref.~\cite{Soumagnac:2016bjk,Soumagnac:2018atx,Heinrich:2019sxl}. The effects of CIPs on 21-cm fluctuations were considered in Ref.~\cite{Gordon:2009wx}, and their implications for the velocity acoustic oscillations \cite{Munoz:2019rhi,Munoz:2019fkt} in the 21-cm power spectrum are discussed in Ref.~\cite{Hotinli:2021xln}. Finally, Refs.~\cite{Barreira:2019qdl,Barreira:2020lva} assessed the sensitivity of galaxy clustering to the amplitude of CIPs through the measurement of the scale-dependent galaxy bias induced due to CIPs.

Here, we study the prospects to use kinetic Sunyaev-Zeldovich (kSZ) tomography to seek uncorrelated CIPs. kSZ tomography \cite{Zhang:2000wf, Ho:2009iw, Shao:2016uzt, Zhang:2010fa, Munshi:2015anr,Smith:2018bpn, Cayuso:2021ljq} allows for the reconstruction of the line-of-sight component of the peculiar-velocity field in a 3-dimensional volume. This is accomplished by cross-correlating the peculiar velocity-induced temperature fluctuation (the kSZ effect \cite{Sunyaev:1980nv, Zeldovich:1969ff, Zeldovich:1969sb, Sunyaev:1980vz, Sazonov:1999zp}), in a CMB map, with a large-scale galaxy survey, allowing for a measurement of the kSZ contribution as a function of redshift.  Given that the total-matter field can be reconstructed from the velocity field, kSZ tomography provides the ideal arena for testing models, like those with CIPs, in which baryons and dark matter may be set apart from each other. In recent works \cite{Hotinli:2019wdp, Sato-Polito:2020cil}, the improvement coming from the addition of this independent tracer was explored for models of correlated CIPs in which the CIP is fully correlated with the adiabatic perturbation.

The above methodology allows us to compare the kSZ tomography-based matter reconstruction field to galaxy survey data to obtain excellent constraints on the amplitude of the CIP power spectrum. In fact, because these two tracers of the large-scale matter distribution are obtained independently, we can construct an estimator that compares the amplitude of the galaxy-density fluctuation with that of the matter-density fluctuation for each Fourier amplitude. This estimator is thus not cosmic-variance limited and can, in principle (in the limit of perfect measurements), probe an arbitrarily small CIP amplitude.  Given that the estimator works on a mode-by-mode basis, it also works for correlated CIPs, although it does not capitalize upon additional effects induced by the correlation~\cite{Hotinli:2019wdp}. 

In this paper, we explain the construction of this estimator and make forecasts on the sensitivity of kSZ tomography to the CIP power-spectrum amplitude $A$. We construct the estimator assuming that the CIP is a primordial perturbation field. We note that the CIP amplitude is, strictly speaking, degenerate with a CIP bias that relates the CIP perturbation to the galaxy-density perturbation it induces.  This CIP bias is, however, expected to be of order unity and can be obtained from simulations \cite{Barreira:2019qdl,Barreira:2020lva}. Furthermore, in the event that the effects of CIPs are detected in the CMB, their effects in kSZ tomography can then be used to establish the CIP bias.

In our forecasts, we consider two baseline experiment configurations: `baseline 1' matching the expected specifications of the Vera Rubin Observatory (VRO) \cite{LSSTScience:2009jmu} and CMB-S4 \cite{CMB-S4:2016ple}, and `baseline 2' corresponding to the Dark Energy Spectroscopic Instrument (DESI) \cite{DESI:2016fyo} and Simons Observatory (SO) \cite{SimonsObservatory:2018koc, SimonsObservatory:2019qwx}. We find that baseline 1 results in a sensitivity of $\sigma_{\hat{A}} \approx 2.3 \times 10^{-9}$, where the errors represent the root variance with which the CIP power spectrum amplitude $A$ can be determined. Similarly, we forecast that the expected sensitivity of baseline 2, based on our minimum variance estimator, is $\sigma_{\hat{A}} \approx 5.4 \times 10^{-8}$. These results indicate that it may be possible to probe CIP perturbations with an amplitude comparable to the amplitude of the primordial power spectrum $A_s$. More specifically, we find a relative uncertainty of $\sigma_{\hat{A}} / A_s \approx 1.0$ and $\sigma_{\hat{A}} / A_s \approx 25$ for each of the baselines, respectively, where we use the value of $A_s$ quoted by the \textit{Planck} 2018 CMB analysis \cite{Planck:2018jri}.

This paper is organized as follows:  In Sec. \ref{sec:cip}, we introduce our parameterization of the CIP model, and in Sec. \ref{sec:mve}, we derive the minimum-variance estimator for the CIP amplitude.  We detail the relevant models for the noise and power spectra used in our analysis in Sec. \ref{sec:NoiseModels}. We then present our results in Sec. \ref{sec:results}. For all our analysis we adopt the $\Lambda$CDM Cosmology as the fiducial model with the following parameter values, taken from \textit{Planck} 2018 \cite{Planck:2018jri}: reduced Hubble constant $h = 0.67$, baryon density parameter $\Omega_b = 0.049$, cold dark matter density parameter $\Omega_{\rm{cdm}} = 0.264$, spectral index $n_s = 0.965$ and amplitude of the primordial scalar power spectrum $A_s = 2.2\times10^{-9}$. These forecasts represent a considerable improvement over current constraints $A\lesssim 0.01$ from the CMB \cite{Planck:2018jri,Grin:2013uya} and galaxy clusters \cite{Holder:2009gd,Grin:2013uya}, although should be viewed as complementary to the cluster constraint which probes wave numbers primarily around $k\sim0.1$~Mpc$^{-1}$ as opposed to Hubble scales $k\sim10^{-4}$~Mpc$^{-1}$.

\section{Compensated isocurvature perturbations}
\label{sec:cip}

\subsection{Definitions and conventions}

We define the CIP field $\Delta(\vec x)$ to be the primordial fractional baryon overdensity through
\begin{equation}
     \rho_b(\vec x,z) = \bar\rho_b(z)\left[ 1+\Delta(\vec x) \right],
\end{equation}
which is then accompanied by a compensating dark-matter underdensity,
\begin{equation}
     \rho_c(\vec x,z) = \bar \rho_c(z) \left[ 1 - f_b \Delta(\vec x) \right].
\end{equation}
Here $\bar \rho_b(z)$ and $\bar\rho_c(z)$ are respectively the mean baryon and dark matter densities at redshift $z$, and $f_b$ is the ratio $\Omega_b/ \Omega_c$ today. These defining relations are understood to be valid at sufficiently early times, such that the dark matter and baryons have not moved  significantly, either due to nonlinear evolution at late times or before recombination due to the tight coupling of baryons to photons. Therefore, this setup leads to a modulation of the relative fraction of baryons and dark matter on large scales, while keeping the total matter density fixed.

The CIP perturbation $\Delta(\vec x)$ is a realization of a random field with power spectrum $P_{\Delta\Delta}(k)=A F(k)$, which we have written in terms of an amplitude $A$ and fiducial $k$ dependence $F(k)$. Under the assumption of Gaussian, slow-roll inflation, the canonical choice for the $k$ dependence is the scale-invariant power spectrum $F(k)=1/k^3$ (see, for example, Ref. \cite{Byrnes:2006fr}). This choice is also consistent with the latest \textit{Planck} satellite CMB analysis presented in Ref. \cite{Planck:2018jri}.  In this case, the CIP variance, smoothed in spheres of radius $R$, is \cite{Smith:2017ndr}
\begin{equation}
     \Delta_{\rm rms}^2(R) = \frac{1}{2\pi^2} \int\, k^2\, dk\, \left[ 3 j_1(kR)/(kR) \right]^2 P_{\Delta\Delta}(k),
\end{equation}
where $j_1(x)$ is the spherical Bessel function. If we take $R$ to be the CMB scale considered in Ref.~\cite{Smith:2017ndr}, then $\Delta_{\rm rms}^2 \simeq A/4$.  The current constraints to this amplitude (for uncorrelated CIPs) are $\Delta_{\rm rms}^2 = 0.0037^{+0.0016}_{-0.0021}$ from Planck \cite{Planck:2018jri}, $\Delta_{\rm rms}^2 \lesssim 0.012$ (95\% C.L.) from the WMAP trispectrum \cite{Grin:2013uya}, and $\Delta_{\rm rms}^2 \lesssim 0.006$ from baryon fractions in galaxy clusters \cite{Holder:2009gd,Grin:2013uya}.

\subsection{CIPs and the galaxy perturbation}

Following Ref.~\cite{Barreira:2020lva}, the linear-order expression for the fractional galaxy-density perturbation at comoving position $\bm{x}$ and redshift $z$ can be written
\begin{equation}
    \delta_g(\bm{x},z) = b_g(z) \delta_m(\bm{x},z) +  b_{\rm CIP}(z) \Delta(\bm{x}),
\label{eqn:deltag}    
\end{equation}
where $b_g(z)$ is the usual linear galaxy bias, and $b_{\rm CIP}(z)$ is a CIP bias that parametrizes the contribution of the CIP to the galaxy-density perturbation.  The fractional matter-density perturbation $\delta_m(\bm{x},z)$ is taken to be the large-scale matter perturbation which grows proportional to the linear-theory growth factor. Given that the CIP generates no gravitational-potential perturbation, $\Delta(\bm{x})$ will remain approximately constant on large-distance scales and so has no redshift dependence.  

The relation between $\Delta(\bm{x})$ and $\delta_g(\bm{x},z)$, parameterized by the CIP bias $b_{\rm CIP}(z)$, can be obtained through simulations.  This bias is determined by two competing effects: (1) the effect on the halo mass function, which decreases with increasing $\Delta$; and (2) the ratio of the stellar mass to the halo mass, which increases with $\Delta$.  Simulation results for $b_{\rm CIP}$ depend on whether the galaxies are selected by halo mass or stellar mass. Further details can be found in Refs. \cite{Barreira:2019qdl,Barreira:2020lva}.

Given Eq.~(\ref{eqn:deltag}), the galaxy power spectrum for uncorrelated CIPs will be
\begin{equation}
    P_{gg}(k,z) = b_g^2(z) P_{mm}(k,z) + b_{\rm CIP}^2(z) P_{\Delta\Delta}(k).
    \label{eq: Pgg_Pmm_Pdeltadelta}
\end{equation}
Thus, the CIPs show up as an additional contribution to the galaxy power spectrum.  In principle (and in practice), the CIP contribution $b_{\rm CIP}^2(z) P_{\Delta\Delta}(k)$ to the galaxy power spectrum can be inferred by comparing the observed galaxy power spectrum to the matter power spectrum obtained from the peculiar velocity field determined from kSZ tomography. However, the measurements of both of the power spectra, $P_{gg}(k)$ and $P_{mm}(k)$, are cosmic-variance limited i.e.; they are both independently limited by the number of Fourier modes of the galaxy and velocity fields that can be obtained with high signal to noise. Therefore, using the above model to constrain the CIP amplitude will be limited by the effects of cosmic variance on each of the measured power spectra.

\section{Minimum-variance estimator}
\label{sec:mve}
 With kSZ tomography, the CIP perturbation amplitude can be obtained on a mode-by-mode basis, under (relative) cosmic-variance cancellation. In Fourier space, the estimator for the amplitude $\Delta_{\bm{k}}$ is then
\begin{equation}
    \widehat{\Delta_{\bm{k}}} = \left(\widehat{\delta_{g,\bm{k}}} - b_g \widehat{ \delta_{m,\bm{k}}} \right)/b_{\rm CIP},
\end{equation}
where the overhat denotes an estimator, and we have dropped any redshift dependence for ease of notation. This estimator has a variance (under the null hypothesis $\Delta=0$),
\begin{equation}
    P_{\Delta\Delta}^N(\bm{k}) = \left[ \VEV{\left|\Delta_{\bm{k}} \right|^2} \right]= b_{\rm CIP}^{-2}\left[N_{gg}({\bm{k}}) + b_g^2 N_{mm}({\bm k})\right],
\end{equation}
where $N_{gg}(k)$ and $N_{mm}(k)$ are the noise contributions to the galaxy and matter power spectra, respectively.

The detectability of CIPs can be assessed by determining the error $\sigma_{\widehat A}$ with which the amplitude $A$ for the CIP power spectrum can be measured.  The minimum-variance estimator $\widehat A$ for the amplitude is then obtained by adding the estimators from each Fourier mode with inverse-variance weighting:
\begin{equation}
    \widehat A = b_{\rm CIP}^2 \sigma_{\widehat A}^2 \sum_{\bm{k}} \frac{ \left| \widehat{\delta_{g,\bm{k}}} - b_g \widehat{ \delta_{m,\bm{k}}} \right|^2/F(k)}{ 2 \left[ N_{\Delta\Delta}(\bm{k})/F(k) \right]^2 }.
\end{equation}
Here,
\begin{equation}
    \sigma_{\widehat A}^2 = b_{\rm CIP}^{-4} \left[\frac12 \sum_{\bm{k}} \left[ F(k)/N_{\Delta\Delta}({\bm k}) \right]^2 \right]^{-1},
\label{eqn:variance}    
\end{equation}
is the variance with which the CIP amplitude $A$ can be determined, and we have defined $N_{\Delta\Delta}({\bm k}) \equiv N_{gg}({\bm{k}}) + b_g^2 N_{mm}({\bm k})$ to make explicit the $b_{\rm CIP}$ dependence of the estimator. Since this method relies on measurements of $\widehat{\delta_{g,\bm{k}}}$ and $\widehat{\delta_{m,\bm{k}}}$, we no longer have two independent terms carrying the cosmic-variance limitations. Therefore, using this estimator method, we can decrease the effects of sample variance and increase sensitivity to the CIP amplitude, in comparison to the methodology presented below Eq.~\eqref{eq: Pgg_Pmm_Pdeltadelta}.

\section{Noise Models}
\label{sec:NoiseModels}
We model the noise in the galaxy autopower spectrum assuming that the primary contribution comes from galaxy shot noise along with photo-$z$ errors. Photo-$z$ errors can be implemented by a convolution of the galaxy density field with a Gaussian kernel in the radial direction. The galaxy noise power spectrum is then given by:
\begin{equation}
    N_{gg}(k, \mu) = \frac{1}{W^2(k, \mu) n_{\rm gal}},
\end{equation}
where $n_{\rm gal}$ is the average galaxy number density of the specific survey, and Gaussian kernel $W(k, \mu)$ is defined as
\begin{equation}
    W^2(k, \mu) = e^{-k^2\mu^2\sigma^2(z)/H^2(z)},
\end{equation}
with redshift scattering $\sigma(z)$. 

The noise in the independently-calculated matter-overdensity field is derived from the kSZ velocity reconstruction noise. As shown in Ref. \cite{Smith:2018bpn}, the noise in the kSZ-tomography-based reconstruction of the velocity field is given by
\begin{equation}
    N_{vv}(k_L, \mu_L) = \mu_{L}^{-2}\frac{2\pi \chi_*^2}{K_*^2}\Bigg[\int dk_S \frac{k_SP_{ge}^{\rm NL}(k_S)^2}{P_{gg}^{\rm NL}(k_S)\ C_{\ell=k_S\chi_*}^{\text{tot}}}\Bigg]^{-1},
    \label{eq:NvvFullExpression}
\end{equation}
where $\chi_{*}$ refers to the comoving distance to the redshift of consideration $z_{*}$, $k_L$ refers to the long-wavelength mode, $k_S$ refers to the short-wavelength mode, and $\mu_L$ refers to the angle of the large-scale mode with respect to the line of sight, i.e., $\mu_L = \hat{\bm{k}}_L\cdot \hat{\bm{n}}$. Furthermore, $P_{gg}^{\rm NL}(k_{S}, \mu_{S})$ refers to the small-scale galaxy-galaxy autopower spectrum and $P_{ge}^{\rm NL}(k_{S}, \mu_{S})$ is the small-scale galaxy-electron power spectrum. Finally, in the above equation we use the radial weight function $K_{*}$ given by
\begin{equation}
    K_* \equiv -T_{\text{CMB}}\sigma_T\bar{n}_{e,0}e^{-\tau(\chi_*)}(1+z_*)^2,
\end{equation}
where $\bar{n}_{e,0}$ is the mean electron density today, and $\tau$ is the optical depth. It is important to note that the velocity reconstruction noise is independent of the magnitude of $k_L$. 

Using the late-time, linearized, continuity-equation-based relation between the peculiar-velocity field and matter-overdensity field, we can write the noise in the matter reconstruction as
\begin{equation}
    N_{mm}(k_L, \mu) = \frac{k_L^2}{(faH)_*^2}N_{vv}(k_L, \mu),
    \label{eq:MatReconNvv}
\end{equation}
where $f$ refers to the linear growth rate $d \ln{G}/ d \ln{a}$, $H$ is the Hubble parameter, and $a$ is the scale factor at the redshift of interest. Since $N_{vv}$ is independent of the magnitude of $k_{L}$, the above relation implies that the noise in the reconstructed matter power spectrum is proportional to $k_{L}$; i.e., the noise is lowest on the largest scales. 

The small-scale galaxy-galaxy and galaxy-electron power spectra appearing in Eq.~\eqref{eq:NvvFullExpression} are calculated within the halo model including the halo occupation distribution (HOD) \cite{Leauthaud:2011rj, Leauthaud:2011zt}. The specific modeling assumptions and parameter values used to construct the small-scale spectra can be found in Appendix A of Ref. \cite{AnilKumar:2022flx}. To ensure that the computed small-scale spectra under the HOD model are consistent with the assumed experiment specifications, we use the following prescription. In the HOD model, the galaxy sample is specified by imposing a particular threshold stellar mass $m_\star^{\rm thresh}$ of observable galaxies. For each configuration, we choose an $m_\star^{\rm thresh}$ such that the total predicted number density of observed galaxies matches the number density for the given experiment.

Finally, in order to complete the model of the velocity-reconstruction noise, we define the CMB contribution as follows. The total CMB contribution $C_\ell^{\rm tot}$, appearing in Eq.~\eqref{eq:NvvFullExpression}, is assumed to be
\begin{equation}
    C_{\ell}^{\text{tot}} = C_{\ell}^{TT} + C_{\ell}^{\text{kSZ-late-time}} + N_{\ell},
    \label{eq:Cll_contributions}
\end{equation}
where $C_{\ell}^{TT}$ is the lensed CMB temperature power spectrum, $C_{\ell}^{\text{kSZ-late-time}}$  is the low-redshift contribution to kSZ, and finally, $N_{\ell}$ is the instrumental-noise power spectrum of the CMB map, which is modeled as
\begin{equation}
    N(\ell) = s^2\text{exp}\Bigg[\frac{\ell(\ell + 1)\theta_{\rm FWHM}^2}{8\ \text{ln}2}\Bigg].
    \label{eq:CMBNoise}
\end{equation}
Here, $s$ labels the sensitivity of the instrument, and $\theta_{\rm FWHM}$ is the resolution. We do not include a contribution from atmospheric noise since it is expected to be subdominant to the instrument and kSZ contributions at the relevant high multipoles of $\ell > 3000$.

\section{Results}
\label{sec:results}

In this section, we provide forecasts for two different experimental configurations, choosing a fixed, fiducial set of values for the survey parameters to model the noise expected in each case. We then present the dependence of $\sigma_{\hat{A}}$ on the survey parameters by varying each independently, to better establish the direction for improvements to future surveys. 

\subsection{Baseline forecasts}

\begin{figure*}
    \centering
    \includegraphics[width=\columnwidth]{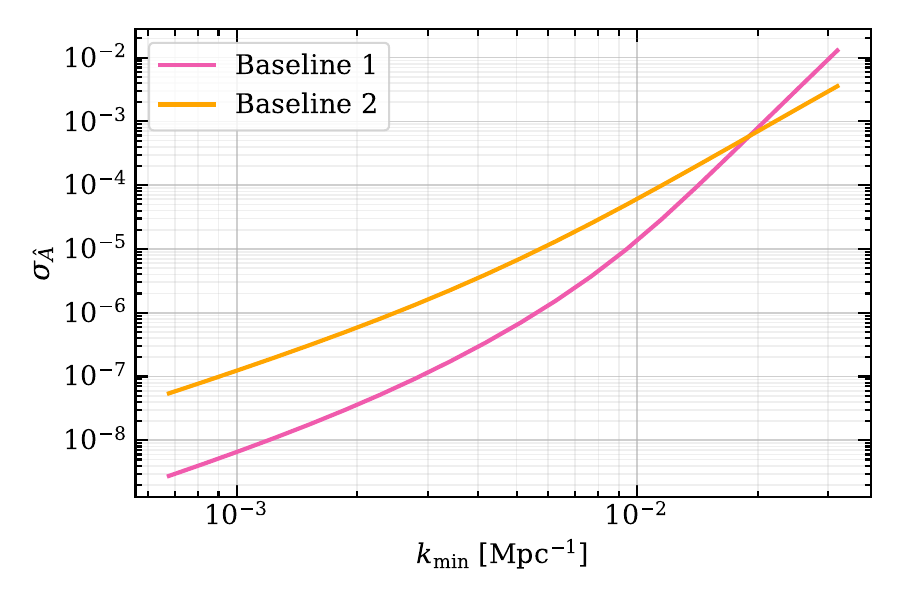}%
    \centering
    \includegraphics[width=\columnwidth]{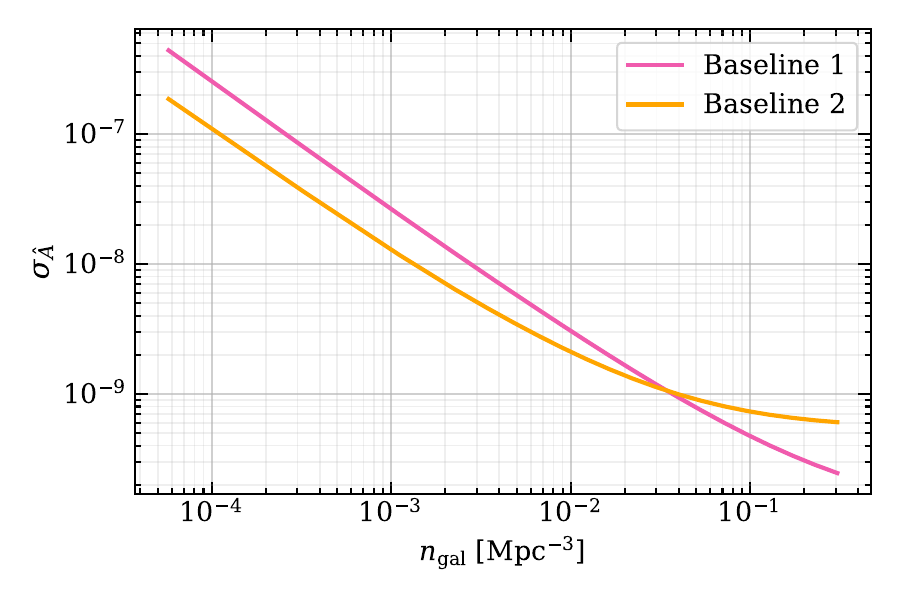}%
    \caption{\textit{Left:} $\sigma_{\hat{A}}$ as a function of $k_\text{min}$ for each of the baselines. The lower cutoff for $k_{\rm min}$, on the left, is defined by the survey volume $V$. For lower $k_{\rm min}$ baseline 1 performs better due to VROs higher $n_{\rm gal}$. For larger values of $k_{\rm min}$, photo-$z$ errors dominate the noise in baseline 1, and baseline 2 performs better, in comparison. \textit{Right:} $\sigma_{\hat{A}}$ as a function of $n_{\rm gal}$ for each of the baselines. The significant increase in sensitivity to the CIP power-spectrum amplitude $\hat{A}$ with increasing galaxy number density occurs to the lowered shot noise which allows for a better reconstruction of the large-scale galaxy and matter overdensity modes. Baseline 2 performs better at lower $n_{\rm gal}$ due to spectroscopic redshift measurements, however, the increased CMB noise for this configuration causes results to plateau at larger $n_{\rm gal}$.}%
    \label{fig:error_dep_kmin_ngal}%
\end{figure*}

We forecast future sensitivity to the amplitude of the CIP by evaluating Eq~\eqref{eqn:variance} for two experimental configurations: (1) a high galaxy-number density, photometric survey similar to VRO \cite{LSSTScience:2009jmu} along with a CMB experiment with specifications that match CMB-S4 \cite{CMB-S4:2016ple}, and (2) a low galaxy number density, spectroscopic survey like DESI \cite{DESI:2016fyo} with a CMB experiment like SO \cite{SimonsObservatory:2018koc}. The set of experimental survey parameters used in our calculation have been taken from Ref. \cite{AnilKumar:2022flx}, and are summarized in Table \ref{tab:baselineSpecs}. 

\begin{table}
\caption{\label{tab:baselineSpecs}%
Baseline configurations for the cross-correlated CMB and LSS experiments. Values for baseline 1 match the specifications of the VRO survey and CMB-S4. The values for baseline 2 are similar to those expected for DESI and SO. The chosen values for the CIP bias are taken from Table 1 of Ref. \cite{Barreira:2020lva}. The survey volumes are the same across the two configurations to emphasize the dependence of the results on galaxy number density and photo-$z$ errors.
}
\begin{ruledtabular}
\begin{tabular}{cccc}
& & \textrm{baseline 1} & \textrm{baseline 2}\\
\colrule
redshift & $z$ & 1.0 & 1.0\\
survey volume & $V$ & 100 $\ \text{Gpc}^{3}$ & 100$\ \text{Gpc}^3$\\
halo bias & $b_h$ & 1.6 & 1.6\\
galaxy density & $n_{\rm{gal}}$ & $10^{-2}\ \ \text{Mpc}^{-3}$ & $2\times10^{-4}\ \ \text{Mpc}^{-3}$\\
photo-$z$ error & $\sigma_z$ & 0.06 & - \\
threshold mass & $m_\star^{\rm thresh}$ & $10^{9.5}\ M_\odot$ & $10^{11}\ M_\odot$\\
CIP bias & $b_{\rm CIP}$ & 0.32 & 0.40\\
CMB resolution & $\theta_{\rm FWHM}$ & 1.5 arcmin & 1.5 arcmin\\
CMB sensitivity & $s$ & 1 $\ \mu \rm{K}-$arcmin & 5 $\ \mu \rm{K}-$arcmin\\
\end{tabular}
\end{ruledtabular}
\end{table} 

It is important to note that the CIP bias $b_{\rm CIP}$ is degenerate with the CIP amplitude. Despite this degeneracy, in our constructed estimator $\hat{A}$, and the associated variance $\sigma_{\hat{A}}$, we continue to treat $A$ and $b_{\rm CIP}$ as separate parameters to clearly establish the dependence of $\sigma_{\hat{A}}$ on the chosen value of the bias. The exact value $b_{\rm CIP}$ can be computed using simulations, and is expected to be of order unity, as presented in Refs. \cite{Barreira:2019qdl,Barreira:2020lva}. To remain consistent with our previous definitions, for these forecasts, we fix the value of $b_{\rm CIP}$, assuming that the galaxy samples are selected by a threshold stellar mass $m^{\rm thresh}_{\star}$. The $m^{\rm thresh}_{\star}$ values are chosen to match the predicted galaxy number density of each survey and are consistent with the small-scale galaxy power spectra used to compute the velocity reconstruction noise for each experimental configuration. The assumed value of $m_\star^{\rm thresh}$ for each survey along with the corresponding value of $b_{\rm CIP}$, estimated from the results in \cite{Barreira:2019qdl,Barreira:2020lva}, have also been included in Table \ref{tab:baselineSpecs}. 

Instead of discretely summing over the Fourier modes, to compute $\sigma_{\widehat A}$, we evaluate Eq.~\eqref{eqn:variance} in the continuous limit as follows:
\begin{equation}
\begin{split}
    \sigma_{\widehat A}^2 &= b_{\rm CIP}^{-4}\left[\frac{V}{2}\int \frac{dk^3}{(2\pi)^3}\left(\frac{F(k)}{N_{\Delta\Delta}(\bm{k})}\right)^2\right]^{-1}, \\
    &=  b_{\rm CIP}^{-4}\left[\frac{V}{2}\int_{k_{\rm min}}^{k_{\rm max}} \int_{-1}^{1} \frac{k^2dk\ d\mu}{(2\pi)^2}\left(\frac{F(k)}{N_{\Delta\Delta}(\bm{k})}\right)^2\right]^{-1},
\end{split}
\label{eq: sigmaA_integral}
\end{equation}
where we have accounted for the fact that the variance $N_{\Delta\Delta}$ is only dependent on $k$ and $\mu$, with the latter being induced by the kSZ-based velocity reconstruction and the inclusion of photo-$z$ errors. The value of $b_g$ is completely defined by our choices for $n_{\rm{gal}}$ and halo bias $b_h$, given the HOD model specifications from Ref. \cite{AnilKumar:2022flx}. Furthermore, on large scales, where we expect the signal to be dominant, we can approximate $b_g \approx b_h$. For our forecasts, we adopt the canonical choice $F(k) = 1/k^3$. The integral over Fourier modes is performed from a lower limit $k_\text{min} \equiv \pi/V^{1/3}$, restricted by the survey volume $V$, to an upper limit $k_\text{max} \approx 10^{-1} \ \text{Mpc}^{-1}$.

Through our analysis, we find that for the configuration of VRO and CMB-S4, $\sigma_{\hat{A}} \approx 2.3\times10^{-9}$ which corresponds to a relative sensitivity of $\sigma_{\hat{A}}/A_s \approx 1.0$, where $A_s$ is the amplitude of the primordial power spectrum. Similarly, for the configuration of DESI and an SO-like CMB experiment, we find that $\sigma_{\hat{A}} \approx 5.4 \times 10^{-8}$ with a relative uncertainty of $\sigma_{\hat{A}}/ A_s \approx 25$. For these relative uncertainty estimates, we use the value $A_s = 2.2 \times 10^{-9}$ determined by the most recent \textit{Planck} 2018 CMB analysis \cite{Planck:2018jri}. 

\subsection{Experiment parameter variations}

In order to assess which experimental limitations have the most significant impact on our ability to measure the CIP power spectrum amplitude, we isolate the effects of certain experimental parameters from Table \ref{tab:baselineSpecs} by varying each individually and holding all other elements of the configuration constant. The results of these variations are discussed below. 

First, to highlight the scales that most prominently contribute to the signal, we plot the value of $\sigma_{\hat{A}}$ as a function of the smallest measurable Fourier mode $k_{\rm min}$. This variation corresponds to changing the largest recoverable wave number from (fixed) survey volume $V$, and directly impacts the lower limit of `summation' evaluated via Eq.~\eqref{eq: sigmaA_integral}. The results for both baselines have been presented in Fig. \ref{fig:error_dep_kmin_ngal} (left). The displayed results indicate that the inclusion of larger scales increases survey sensitivity to the CIP power spectrum amplitude. This is an expected result, not only because the CIP signal is largest at small $k$ [since we have chosen $F(k)\sim 1/k^3$ for this analysis], but also because the noise in the reconstructed matter overdensity field is smallest on largest scales [see Eq.~\eqref{eq:MatReconNvv}]. At lower values of $k_{\rm min}$, baseline 1 performs better than baseline 2, likely due to the lowered shot noise (higher $n_{\rm gal}$). However, baseline 2 performs better at higher $k_{\rm min}$, where the effects of shot noise are minimized and the photo-$z$ errors become dominant in the baseline 1 estimates.

Next, we focus on highlighting the effects of increasing galaxy number density $n_{\rm gal}$ on the value of $\sigma_{\hat{A}}$. The results for this variation for each of the baselines (holding all other experimental parameters constant, for each individual baseline) can be seen in Fig. \ref{fig:error_dep_kmin_ngal} (right). The results displayed show that an increasing galaxy number density allows for higher survey sensitivity to the CIP power spectrum amplitude $A$. This behavior is a direct result of the fact that a higher galaxy number density equates to a lower shot noise, which not only allows for the measurement of the larger-scale galaxy modes but also decreases the matter reconstruction noise (through a lower overall $N_{vv}$). The two curves are relatively parallel for $10^{-4}\ {\rm Mpc}^{-3} < n_{\rm gal} < 10^{-2}\ {\rm Mpc}^{-3}$, with baseline 2 performing better in this region due to spectroscopic redshift measurements. However, baseline 1 (VRO+CMB-S4) performs better at a higher galaxy number density, while the results from baseline 2 (DESI+CMB-SO) plateau, likely due to the difference in CMB resolutions.

\begin{figure}
\includegraphics[scale=0.55]{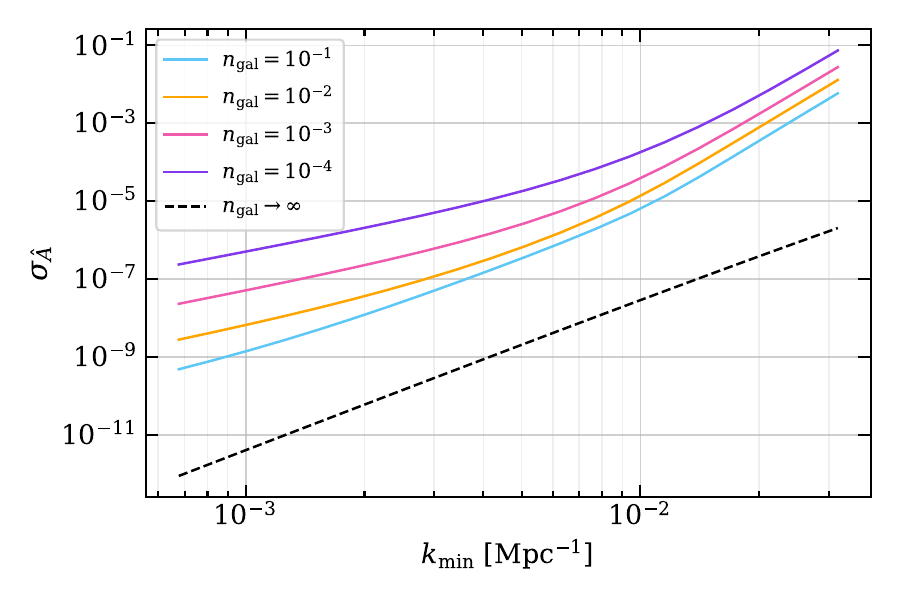} 
\caption{$\sigma_{\hat{A}}$ as a function of $k_\text{min}$ for various values of $n_{\rm gal}$ (in ${\rm Mpc}^{-3}$). The results indicate that an increasing galaxy number density improves the steepness of decrease in $\sigma_{\hat{A}}$ with decreasing $k_{\rm min}$, most evidently for $10^{-3}\ {\rm Mpc}^{-1} < k_{\rm min} < 10^{-2}\ {\rm Mpc}^{-1}$. The dashed line displays the `ideal' case in which $n_{\rm gal}$ is infinitely large i.e., the shot noise is zero.}
\label{fig:fig:error_dep_kmin_diff_ngal}
\end{figure}

What is more interesting to analyze is the effect of galaxy number density on the relation between $\sigma_{\hat{A}}$ and $k_{\rm min}$. Figure \ref{fig:fig:error_dep_kmin_diff_ngal} displays $\sigma_{\hat{A}}$ as a function of $k_{\rm min}$ for different values of $n_{\rm gal}$. For these curves, we assume the baseline 1 configuration for all other survey parameters and keep the value of $b_{\rm CIP}$ fixed. The displayed results indicate that a higher galaxy number density results in a steeper decrease of $\sigma_{\hat A}$ with decreasing $k_{\rm min}$ i.e., a higher $n_{\rm gal}$ allows for a greater order-of-magnitude improvement in $\sigma_{\hat A}$ with a fixed increase in survey volume. This effect is particularly evident for $10^{-3}\ {\rm Mpc}^{-1} < k_{\rm min} < 10^{-2}\ {\rm Mpc}^{-1}$. The black dashed line, labeled `No Noise', portrays the dependence of $\sigma_{\hat{A}}$ on $k_{\rm min}$ in the absence of shot noise ($n_{\rm gal} \rightarrow \infty$) and photo-$z$ errors. In this ideal case, we see that $\sigma_{\hat{A}}$ approximately scales as $k_{\rm min}^{3.5}$. This behavior is explained by the chosen model for $P_{\Delta\Delta}(k)$ [with $F(k) = 1/k^3$] along with the $k^2$ scale dependence of the matter reconstruction noise [Eq.~\eqref{eq:MatReconNvv}].

On the contrary, assuming the same baseline configuration as above, we found that varying $k_{\rm min}$ between $10^{-3}\ {\rm Mpc}^{-1}$ and $10^{-2}\ {\rm Mpc}^{-1}$ has a minimal impact on the steepness of the dependence of $\sigma_{\hat{A}}$ on the galaxy number density. That is, even though a decreased $k_{\rm min}$ improves sensitivity to the CIP power-spectrum amplitude, a fixed increase in the galaxy number density consistently leads to a fixed order-of-magnitude improvement in $\sigma_{\hat {A}}$ for $10^{-3}\ {\rm Mpc}^{-1} < k_{\rm min} < 10^{-2}\ {\rm Mpc}^{-1}$. The results only significantly diverge for $n_{\rm gal} > 10^{-1.75}\ {\rm Mpc}^{-3}$, which is likely due to the shot noise becoming sub-dominant at these higher values of galaxy number density.

Finally, to highlight the effects of CMB noise on survey sensitivity to $A$, for each of the discussed baseline configurations, we varied the CMB telescope sensitivity $s$, and resolution $\theta_{\rm FWHM}$ individually, holding all other experimental parameters constant. We found that, once again, the difference in galaxy number density across the two baselines severely impacts the order of magnitude improvement in $\sigma_{\hat{A}}$, given a fixed improvement in CMB noise parameters. We vary the CMB sensitivity from $0.25\ \mu$K-arcmin to $10\ \mu$K-arcmin and find a steady increase in $\sigma_{\hat{A}}$ by a factor of 3 for the baseline 1 configuration and a factor of 1.1 for the baseline 2 configuration. Similarly, we vary CMB telescope resolution from 0.1 arcmin to 10 arcmin and find a relatively steady increase in both cases, by a factor of 250 for the baseline 1 configuration and a factor of 40 for baseline 2. This indicates that, at a higher value of $n_{\rm gal}$, surveys are more sensitive to increases in CMB instrument noise. 

For completeness, we also varied the photo-$z$ error assumed for baseline 1, holding all other experiment parameters fixed. Varying the value of $\sigma_z$ from 0.0 to 2.0 resulted in an increase in $\sigma_{\hat A}$ by a factor of 3.5. This minimal effect from increasing photo-$z$ is expected, given that we are primarily reliant on the signal from the largest scales for the measurement. We also varied the assumed value of the Gaussian galaxy bias $b_g$ for both the baseline configurations (holding $n_{\rm{gal}}$ fixed) to conclude that its effect on survey sensitivity to $\sigma_{\hat{A}}$ is negligible.

\section{Discussion}

In this paper, we forecast future survey sensitivity to the amplitude of the CIP power spectrum $A$, assuming that the compensated perturbations are sourced primordially. The compensated nature of these isocurvature perturbations causes CIPs to contribute only at second order to the CMB, leading to poor constraints that allow for the CIP amplitude to be over five orders of magnitude larger than that of the primordial adiabatic perturbation. In contrast, the CIP amplitude is expected to contribute at leading order to the galaxy overdensity field [see Eq.~\eqref{eqn:deltag}], making it a valuable statistic to investigate the CIP amplitude. Therefore, in our work, we construct a minimum variance estimator that compares the amplitude of the galaxy density fluctuation to the independently obtained matter overdensity amplitude, on a mode-by-mode basis. We show that leveraging the ability to measure the matter over-density field using kSZ tomography, independently of the galaxy over-density field, allows one to probe CIP amplitudes as small as that of the primordial adiabatic perturbation, under sample variance cancellation.

We use the minimum-variance estimator to forecast that a survey configuration corresponding to CMB-S4 and VRO results in a sensitivity of $\sigma_{\hat{A}} \approx 2.3 \times 10^{-9}$. Similarly, a configuration corresponding to SO and DESI results in a sensitivity of $\sigma_{\hat{A}} \approx 5.4 \times 10^{-8}$. These sensitivities correspond to relative uncertainties of $\sigma_{\hat{A}}/A_s \approx 1.0$ and $\sigma_{\hat{A}}/A_s \approx 25$ for each of the combinations, respectively, where $A_s$ represents the amplitude of the primordial power spectrum. For these forecasts, we assume a fixed value for the CIP bias $b_{\rm CIP}$ for each configuration, drawing from the simulation-based results presented in Refs. \cite{Barreira:2019qdl, Barreira:2020lva}. Although the CIP bias is, strictly speaking, perfectly degenerate with the CIP perturbation amplitude, we choose not to consolidate these two parameters into a single amplitude term to make explicit the dependence of $\sigma_{\hat{A}}$ on the value of $b_{\rm CIP}$. Furthermore, since this dependence is just a factor of scale, it is straightforward to map the sensitivities quoted in this paper to a different value of $b_{\rm CIP}$ or to a constraint on a consolidated amplitude parameter $b_{\rm CIP}^{2} \times A$.

The dramatic improvement in sensitivity to CIPs derives from the possibility, enabled by kSZ tomography, to measure the galaxy and total-matter fields independently and thereby circumvent the cosmic-variance limit in many other probes.  Thus, even one very well-measured Fourier mode allows the CIP to be probed.  Our results indicate, moreover, that the sensitivity comes primarily from measurements at the largest scales, a consequence largely of the $k$ dependence of the relation between the total-matter perturbation and the peculiar velocity probed by the kSZ effect.  We thus conclude that in order for the promising statistical errors forecast here to be achieved, systematic effects that might affect the measurement of the galaxy-density field and CMB-temperature perturbations on the largest distance scales must be well under control.  We also surmise that relativistic effects will need to be included in the analysis.

The sensitivity to the CIP amplitude we forecast here compares well (within a factor of $\sim4$) with Ref.~\citep{Hotinli:2019wdp}, where authors evaluated the prospects to probe \textit{correlated} CIP fluctuations with kSZ tomography. Most recent upper limits on CIPs amplitude are provided by the scale-dependent mass-to-light ratio from measurements of BAOs~\citep{2016PhRvL.116t1302S,Soumagnac:2018atx}, which are comparable to the constraints from the CMB~\citep{Smith:2017ndr}, of the order $\sigma_A\sim\mathcal{O}(10^{-4})$. These constraints also compare well with forecasted sensitivities on the BAO phase shift, induced by spatially varying correlated CIP fluctuations, explored in Ref.~\citep{Heinrich:2019sxl}. More recently, Ref.~\citep{Hotinli:2021xln} proposed using measurements of the velocity acoustic oscillations (VAOs) during cosmic dawn~\citep{Munoz:2019rhi,Munoz:2019fkt} to probe both correlated and uncorrelated CIPs fluctuations at a sensitivity reaching $\sigma_A\sim\mathcal{O}(10^{-5})$ in the foreseeable future. These studies find that the sensitivity of the kSZ tomography studied here and in Ref.~\citep{Hotinli:2019wdp} will likely remain orders of magnitude better compared to that of CMB, BAO, and the VAO signals. 

Constraining the CIP amplitude at higher order will not only allow for a better understanding of whether baryon and CDM fluctuations trace the matter density but also will help rule out different, nontrivial models of many-field inflation. In fact, to accurately probe signatures of deviations from adiabaticity and Gaussianity of the early Universe, accounting for CIPs may be essential. For example, Ref.~\citep{Barreira:2020lva} shows that, depending on the degree of correlation of the CIP with the primordial adiabatic perturbation, the CIP signal may exactly match the scale-dependent signal from the $f_{\rm NL}$ term when probing scale-dependent bias for signatures of primordial non-Gaussianity in the single field scenario. In the curvaton scenario, depending on the correlation coefficient assumed between the CIP and the inflaton or the curvaton, we would expect similar degeneracies to arise when using the galaxy bias to simultaneously probe $f_{\rm NL}$ and $\tau_{\rm NL}$. Such degeneracies may also affect the fidelity of lensing data extracted from the CMB, due to similarities between the effects of lensing and CIPs on the CMB two-point statistics~\citep{Heinrich:2016gqe}.

This emphasizes the importance of considering CIPs to make unbiased measurements of early Universe characteristics. Although we do not consider the effects of non-Gaussianities in our current estimator construction and make a simple set of forecasts under the null hypothesis, we highlight the effectiveness of kSZ tomography as a probe for early universe cosmology. When considering more complicated models including the CIP, we expect cross-correlation tools such as the kSZ tomography, multi-tracer analysis with different populations of galaxies and haloes, CMB lensing, and many others to be essential in obtaining tighter constraints under sample variance cancellation and breaking degeneracies across the varying signatures of the inflationary Universe.

\begin{acknowledgments}
We would like to thank Gabriela Sato-Polito for helpful discussions. This work was supported by NSF Grant No.\ 2112699 and the Simons Foundation. S.C.H. is supported by the Horizon Fellowship at Johns Hopkins University.
\end{acknowledgments}

\bibliography{kSZCIP}
\end{document}